\documentclass[aps,pra,showpacs,superscriptaddress,preprint]{revtex4}

\begin{document}
\title{{\bf Exact Polynomial Eigensolutions of the Schr\"{o}dinger Equation for the
Pseudoharmonic Potential }\\
Sameer M. Ikhdair\thanks{%
sikhdair@neu.edu.tr} and \ Ramazan Sever\thanks{%
sever@metu.edu.tr}}
\address{$^{\ast }$Department of Physics, Near East University, Nicosia, North
Cyprus, Mersin 10, Turkey.\\
$^{\dagger }$Department of Physics, Middle East Technical University, 06531
Ankara, Turkey.}
\date{\today}
\author{}

\begin{abstract}
The polynomial solution of the Schr\"{o}dinger equation for the
Pseudoharmonic potential is found for any arbitrary angular momentum
$l$. The exact bound-state energy eigenvalues and the corresponding
eigen functions are analytically calculated. The energy states for
several diatomic molecular systems are calculated numerically for
various principal and angular quantum numbers. By using a proper
transformation, this problem can be also solved very simply using
the known eigensolutions of anharmonic oscillator
potential.{\normalsize \newline}

Keywords{\normalsize :} Pseudoharmonic potential, anharmonic oscillator
potential, Schr\"{o}dinger equation, diatomic molecules, eigenvalues and
eigenfunctions.
\end{abstract}

\maketitle

\begin{verbatim}

\end{verbatim}

\section{Introduction}

\noindent The three-dimensional $(3D)$ anharmonic oscillators are of great
importance in different physical phenomena with many applications in
molecular physics [1]. The solutions of the Schr\"{o}dinger equation for any
$l$-state for such potentials are also of much concern. Morse potential is
commonly used for anharmonic oscillator. However, its wavefunction is not
vanishing at the origin. On the other hand, the Mie-type and also the
Pseudoharmonic potentials do vanish. The Mie-type potential has the general
features of the true interaction energy [1], interatomic and inter-molecular
and dynamical properties in solid-state physics [2]. The Pseudoharmonic
potential may be used for the energy spectrum of linear and non-linear
systems [3]. The Pseudoharmonic and Mie-type potentials [3,4] are two
exactly solvable potentials other than the Coulombic and anharmonic
oscillator.

The anharmonic oscillator and H-atom (Coulombic) problems have been
thoroughly studied in $N$-dimensional space quantum mechanics for any
angular momentum $l.$ These two problems are related together and hence the
resulting second-order differential equation has the normalized orthogonal
polynomial function solution (cf. Ref.[5] and the references therein).

In this brief letter we will follow parallel solution to Refs.[6,7,8] and
give a complete normalized polynomial solution of $3D$ Schr\"{o}dinger
equation with Pseudoharmonic potential, anharmonic oscillator like potential
with an additional centrifugal potential barrier, for any arbitrary $l$%
-state. Further, by a proper transformation, we obtain the eigensolutions of
this problem from the well-known eigensolutions of the anharmonic oscillator
potential. As an application, we present some numerical results of the
energy states of $N_{2},$ $CO,$ $NO$ and $CH$ molecules [9].

The contents of this paper is as follows. In Section II, we give the
eigensolutions of the $3D$ Schr\"{o}dinger equation with Pseudoharmonic
potential and calculate numerically the energy levels for various diatomic
molecular systems. We also obtain the eigensolutions of the Pseudoharmonic
potential from a known anharmonic eigensolutions using a proper
transformation. Finally, in Section III, we give our results and conclusions.

\section{SCHR\"{O}DINGER EQUATION WITH PSEUDOHARMONIC POTENTIAL}

\noindent We wish to solve the Schr\"{o}dinger equation for a pseudoharmonic
potential [3] given by

\begin{equation}
V(r)=D_{0}\left( \frac{r}{r_{0}}-\frac{r_{0}}{r}\right) ^{2},
\end{equation}
where $D_{0}$ is the dissociation energy between two atoms in a solid and $%
r_{0}$ is the equilibrium intermolecular seperation.

For brevity, we write the radial part of the Schr\"{o}dinger equation as
\begin{equation}
\left[ -\frac{\hbar ^{2}}{2\mu }{\bf \nabla }^{2}+V(r)\right] \psi (r,\theta
,\varphi )=E_{nl}\psi (r,\theta ,\varphi ),
\end{equation}
and employing the transformation $\psi (r,\theta ,\varphi )=\frac{R_{nl}(r)}{%
r}Y_{lm}(\theta ,\varphi ),$ to reduce it into the form [6,7]
\begin{equation}
\left\{ \frac{d^{2}}{dr^{2}}-\frac{l(l+1)}{r^{2}}+\frac{2\mu }{\hbar ^{2}}%
\left[ E_{nl}+2D_{0}-\frac{D_{0}r^{2}}{r_{0}^{2}}-\frac{D_{0}r_{0}^{2}}{r^{2}%
}\right] \right\} R_{nl}(r)=0.
\end{equation}
Furthermore, using the dimensionless abbreviations:

\begin{equation}
\rho =r/r_{0};\text{ }\varepsilon ^{2}=\frac{2\mu r_{0}^{2}}{\hbar ^{2}}%
\left( E_{nl}+2D_{0}\right) ;\text{ }\gamma ^{2}=\frac{2\mu r_{0}^{2}}{\hbar
^{2}}D_{0,}
\end{equation}
gives the following simple form equation

\begin{equation}
\frac{d^{2}R_{nl}(\rho )}{d\rho ^{2}}+\left[ \varepsilon ^{2}-\gamma
^{2}\rho ^{2}-\frac{\gamma ^{2}+l(l+1)}{\rho ^{2}}\right] R_{nl}(\rho )=0.
\end{equation}
The behaviour of the solution at $\rho =0,$ determined by the centrifugal
term and its asymptotic behaviour, determined by the oscillator terms,
suggests us to write:
\begin{equation}
R_{nl}(\rho )=\rho ^{q}\exp (-\frac{\gamma }{2}\rho ^{2})g(\rho ),
\end{equation}
with the numerator of $x^{-2}$ term equal to zero leads to

\begin{equation}
q=\frac{1}{2}\pm \sqrt{\left( l+\frac{1}{2}\right) ^{2}+\gamma ^{2}}.\text{
\ \ }
\end{equation}
As $q>0,$ the above wavefunction vanishes at $\rho =0,$ corresponding to the
strong repulsion between the two atoms. It is reasonable to set Eq.(6) into
Eq.(5) and then to use, instead of $\rho ,$ the variable:
\begin{equation}
s=\gamma \rho ^{2},
\end{equation}
giving the general type of Kummer's (Confluent Hypergeometric) differential
equation

\[
sg^{\prime \prime }(s)+\left[ 1+\sqrt{\left( l+\frac{1}{2}\right)
^{2}+\gamma ^{2}}-s\right] g^{\prime }(s)
\]
\begin{equation}
-\frac{1}{2}\left( 1+\sqrt{\left( l+\frac{1}{2}\right) ^{2}+\gamma ^{2}}-%
\frac{\varepsilon ^{2}}{2\gamma }\right) g(s)=0,
\end{equation}
with the Kummer's function solution:

\[
g(\rho )=C_{11}F_{1}\left( \frac{1}{2}\left( 1+\sqrt{\left( l+\frac{1}{2}%
\right) ^{2}+\gamma ^{2}}-\frac{\varepsilon ^{2}}{2\gamma }\right) ,1+\sqrt{%
\left( l+\frac{1}{2}\right) ^{2}+\gamma ^{2}};\gamma \rho ^{2}\right)
\]

\begin{equation}
+C_{21}F_{1}\left( \frac{1}{2}\left( 1-\sqrt{\left( l+\frac{1}{2}\right)
^{2}+\gamma ^{2}}-\frac{\varepsilon ^{2}}{2\gamma }\right) ,1-\sqrt{\left( l+%
\frac{1}{2}\right) ^{2}+\gamma ^{2}};\gamma \rho ^{2}\right) \rho ^{-2\sqrt{%
\left( l+\frac{1}{2}\right) ^{2}+\gamma ^{2}}}.
\end{equation}
At $\rho =0,$ the second part of the solution so that $C_{2}=0.$ This
clearly differes from the linear oscillator where no boundary condition
exists at origin. A confluent series behaves asymptotically at large
positive values of its argument as

\begin{equation}
_{1}F_{1}(a,c;z)\rightarrow \frac{\Gamma (c)}{\Gamma (a)}\exp (z)z^{a-c},
\end{equation}
leads us to write

\begin{equation}
R_{nl}(\rho )\rightarrow \rho ^{\frac{1}{2}\pm \sqrt{\left( l+\frac{1}{2}%
\right) ^{2}+\gamma ^{2}}}\exp (-\frac{\gamma }{2}\rho ^{2})\exp (\gamma
\rho ^{2})\rho ^{-\left( 1+\sqrt{\left( l+\frac{1}{2}\right) ^{2}+\gamma ^{2}%
}+\frac{\varepsilon ^{2}}{2\gamma }\right) },
\end{equation}
which is exponentially divergent wavefunction. This divergence can be
avoided, in cutting off the series in Eq.(10), by putting the parameter $%
a=-n,$ with $n=0,1,2,...,$ thus transforming the series into a polynomial of
degree $n.$ Hence

\begin{equation}
\frac{1}{2}\left( 1+\sqrt{\left( l+\frac{1}{2}\right) ^{2}+\gamma ^{2}}-%
\frac{\varepsilon ^{2}}{2\gamma }\right) =-n,
\end{equation}
with

\begin{equation}
\frac{\varepsilon ^{2}}{2\gamma }=\frac{r_{0}}{\hbar }\sqrt{\frac{\mu }{%
2D_{0}}}\left( E_{nl}+2D_{0}\right) .
\end{equation}
Thus, solving Eqs.(13) and (14) for the energy eigenvalues gives

\begin{equation}
E_{nl}=-2D_{0}+\frac{\hbar }{r_{0}}\sqrt{\frac{2D_{0}}{\mu }}\left[ 2n+1+%
\sqrt{\frac{2\mu D_{0}r_{0}^{2}}{\hbar ^{2}}+\left( l+\frac{1}{2}\right) ^{2}%
}\right] ,
\end{equation}
and further from Eqs.(6), (10) and (13), we write the final form of the
wavefunction as
\[
\psi (r,\theta ,\varphi )=N_{nl}r^{-\frac{1}{2}+\sqrt{\frac{2\mu
D_{0}r_{0}^{2}}{\hbar ^{2}}+\left( l+\frac{1}{2}\right) ^{2}}}\exp \left( -%
\sqrt{\frac{\mu D_{0}}{2\hbar ^{2}}}\frac{r^{2}}{r_{0}}\right) \times
\]
\begin{equation}
_{1}F_{1}\left( -n,1+\sqrt{\frac{2\mu D_{0}r_{0}^{2}}{\hbar ^{2}}+\left( l+%
\frac{1}{2}\right) ^{2}};\sqrt{\frac{2\mu D_{0}}{\hbar ^{2}}}\frac{r^{2}}{%
r_{0}}\right) Y_{lm}(\theta ,\varphi ),
\end{equation}
where $N_{nl}$ is a normalization constant to be determined from the
normalization condition and $Y_{lm}(\theta ,\varphi )=\sin ^{m}\theta
P_{n}^{(m,m)}(\cos \theta )\exp (\pm im\varphi )$ is the angular part of the
wave function.

On the other hand, for the sake of simplicity, we can immediately obtain the
energy eigenvalues and the corresponding wave functions of the
Pseudoharmonic potential by transforming Eq.(3) to another
Schr\"{o}dinger-like equation with $L(L+1)=l(l+1)+\frac{2\mu D_{0}r_{0}^{2}}{%
\hbar ^{2}},$

\begin{equation}
\left[ \frac{d^{2}}{dr^{2}}-\frac{L(L+1)}{r^{2}}+\frac{2\mu }{\hbar ^{2}}%
\left( E_{nL}^{\prime }-B^{2}r^{2}\right) \right] R_{nL}(r)=0,
\end{equation}
where

\begin{equation}
E_{nL}^{\prime }=E_{nl}+2D_{0},\text{ }B^{2}=\frac{D_{0}}{r_{0}^{2}},\text{
\ and }L=\frac{1}{2}\left[ -1+\sqrt{\left( 2l+1\right) ^{2}+\frac{8\mu
D_{0}r_{0}^{2}}{\hbar ^{2}}}\right] .
\end{equation}
At this point, we should report that Eq. (17) corresponds to the
Schr\"{o}dinger equation of anharmonic oscillator potential, $%
V(r)=B^{2}r^{2},$ with energy levels
\begin{equation}
E_{nL}^{\prime }=\sqrt{\frac{\hbar ^{2}}{2\mu }}B(4n+2L+3),\text{ }%
n=0,1,2,\cdots ,
\end{equation}
and wave functions

\begin{equation}
\psi (r,\theta ,\varphi )=A_{nL}r^{L}\exp \left( -\sqrt{\frac{\mu }{2\hbar
^{2}}}Br^{2}\right) L_{n}^{(L+\frac{1}{2})}(\sqrt{\frac{2\mu }{\hbar ^{2}}}%
Br^{2})\sin ^{m}\theta P_{n}^{(m,m)}(\cos \theta )\exp (\pm im\varphi ),
\end{equation}
where $m=-(n+L+1).$

Finally, in the light of transformation (18), the eigenvalues (15) and the
eigen functions (16) can be easily determined from the traditional formulas
(19) and (20), respectively, with the Laguerre function being expressed in
terms of Kummer's function, that is, $L_{n}^{(\nu )}(z)=_{1}F_{1}(-n,\nu
+1;z)..$

\section{RESULTS AND CONCLUSIONS}

In this work we have studied the analytical solution for a Pseudoharmonic
potential. Considering this potential, the problem is reduced to a harmonic
oscillator potential plus an additional centrifugal potential barrier of
order $1/r^{2}.$ The exact eigensolutions for this particular case have been
obtained, in a similar way as the Hydrogenic solutions [5,8]. We have
calculated the energy eigenvalues and the corresponding wave functions
considering bound-states for any quantum-mechanical system of any angular
momentum $l$ bound by a pseudoharmonic potential. The present results for
the potential parameters $\gamma =0$ reduces to a Harmonic oscillator
solution.

Finally, we calculate the binding energies of the Pseudoharmonic potential
for $N_{2},$ $CO,$ $NO$ and $CH$ diatomic molecules by means of Eq.(15) with
the potential parameter values [9,10].given in Table 1. The explicit values
of the energy for different values of $n$ and $l$ are shown in Table 2.

\acknowledgments
This research was partially supported by the
Scientific and Technological Research Council of Turkey. The authors
wish to thank the referee(s) for the positive and invaluable
suggestions. S.M. Ikhdair wishes to dedicate this work to his family
for their love and assistance.

\bigskip \bigskip \baselineskip= 2\baselineskip
\bigskip
\newpage
\begin{table}[tbp]
\caption{Reduced masses and spectroscopically determined properties of $%
N_{2},$ $CO,$ $NO$ and $CH$ diatomic molecules in the ground electronic
state.}
\begin{tabular}{lllll}
Parameters\tablenotemark[1]\tablenotetext[1]{The parameter values here are
taken from [10].} & $N_{2}$ & $CO$ & $NO$ & $CH$ \\
\tableline$D_{0}$ $(cm^{-1})$ & $96288.03528$ & $87471.42567$ & $64877.06229$
& $31838.08149$ \\
$r_{0}$ $(A^{\circ })$ & $1.0940$ & $1.1282$ & $1.1508$ & $1.1198$ \\
$\mu $ (amu) & $7.00335$ & $6.860586$ & $7.468441$ & $0.929931$%
\end{tabular}
\end{table}

\begin{table}[tbp]
\caption{Calculated energy eigenvalues of the pseudoharmonic potential for $%
N_{2},$ $CO,$ $NO$ and $CH$ diatomic molecules with different values of $n$
and $l$ in $eV.$}
\begin{tabular}{llllll}
State $(n)$ & $l$ & $N_{2}$ & $CO$ & $NO$ & $CH$ \\
\tableline$0$ & $0$ & $0.1091559$ & $0.1019306$ & $0.0824883$ & $0.1686344$
\\
$1$ & $0$ & $0.3273430$ & $0.3056722$ & $0.2473592$ & $0.5050072$ \\
& $1$ & $0.3278417$ & $0.3061508$ & $0.2477817$ & $0.5085903$ \\
$2$ & $0$ & $0.5455302$ & $0.5094137$ & $0.4122301$ & $0.841380$ \\
& $1$ & $0.5460288$ & $0.5098923$ & $0.4126526$ & $0.8449631$ \\
& $2$ & $0.5470260$ & $0.5108495$ & $0.4134977$ & $0.8521246$ \\
$4$ & $0$ & $0.9819045$ & $0.9168969$ & $0.7419718$ & $1.5141255$ \\
& $1$ & $0.9824031$ & $0.9173755$ & $0.7423944$ & $1.5177087$ \\
& $2$ & $0.9834003$ & $0.9183327$ & $0.7432395$ & $1.5248701$ \\
& $3$ & $0.9848961$ & $0.9197684$ & $0.7445070$ & $1.5356002$ \\
& $4$ & $0.9868903$ & $0.9216825$ & $0.7461969$ & $1.5498843$ \\
$5$ & $0$ & $1.2000916$ & $1.1206384$ & $0.9068427$ & $1.8504983$ \\
& $1$ & $1.2005902$ & $1.1211170$ & $0.9072653$ & $1.8540815$ \\
& $2$ & $1.2015875$ & $1.1220742$ & $0.9081104$ & $1.8612429$ \\
& $3$ & $1.2030832$ & $1.1235099$ & $0.9093779$ & $1.8719729$ \\
& $4$ & $1.2050774$ & $1.1254240$ & $0.9110678$ & $1.8862571$ \\
& $5$ & $1.2075699$ & $1.1278165$ & $0.9131799$ & $1.9040761$%
\end{tabular}
\end{table}


\newpage



\begin{thebibliography}{99}
\bibitem{1}  G. C. Maitland, M. Righby, E. B. Smith and W. A. WAkeham,
Intermolcular forces (Oxford Univ. Press, Oxford, 1987).

\bibitem{2}  M. L. Klein and J. A. Vemebles, Rare gas solids, Vol. 1
(academic Press, New York, 1976).

\bibitem{3}  I. I. Goldman and V. D. Krivchenkov, Problems in quantum
mechanics (Pergamon Press, New York, 1961); Y. Weissman and J. Jortner,
Phys. Lett. A 70 (1979) 177; M. L. Seze, Chem. Phys. 87 (1984) 431; M. Sato
and J. Goodisman, Am. J. Phys. 53 (1985) 350; S. Erkoc and R. Sever, Phys.
Rev. A 37 (1988) 2687.

\bibitem{4}  \c{S}. Erko\c{c} and R. Sever, Phys. Rev. D 30, (1984) 2117; \c{%
S}. Erko\c{c} and R. Sever, Phys. Rev. D 33, (1986) 588.

\bibitem{5}  M. M. Neito, Am. J. Phys. 47 (1979) 1067.

\bibitem{6}  S. Fl\"{u}gge, Practical Quantum Mechanics I, (Springer-Verlag
Berlin, 1971).

\bibitem{7}  S. M. Ikhdair and R. Sever, [arXiv:quant-ph/0611065] to be
published in Int. J. Mod. Phys. E.

\bibitem{8}  R. L. Liboff, Introductory Quantum Mechanics, 4th Ed., (Addison
Wesley, 1301 Sansome St., San Francisco, CA 94111, 2003).

\bibitem{9}  C. Berkdemir, A. Berkdemir and J. Han, Chem. Phys. Lett. 417
(2006) 326.

\bibitem{10}  M. Karplus, and R. N. Porter, Atoms and Molecules: An
Introduction For Students of Physical Chemistry, (Benjamin, Menlo Park, CA,
1970).\bigskip
\end{thebibliography}
\end{document}